\documentclass[a4paper]{jpconf}
\usepackage{graphicx}

\begin{document}

\title{Precision Calorimetry at HL-LHC: Upgrade of the CMS Electromagnetic Calorimeter.\\
\emph{Talk presented at the International Workshop on Future Linear Colliders (LCWS2021), 15-18 March 2021}}


\author{S. Argir\`o}




\address{University of Torino, Italy, and INFN}

\ead{stefano.argiro@cern.ch}

\author{on behalf ot the CMS Collaboration}

\begin{abstract}
The electromagnetic ealorimeter (ECAL) of the CMS detector has played an important role in the physics program of the experiment, delivering outstanding performance throughout data taking. The High-Luminosity LHC will pose new challenges. The four to five-fold increase of the number of interactions per bunch crossing will require superior time resolution and noise rejection capabilities. For these reasons the electronics readout has been completely redesigned. A dual gain trans-impedance amplifier and an ASIC providing two 160 MHz ADC channels, gain selection, and data compression will be used in the new readout electronics. The trigger decision will be moved off-detector and will be performed by powerful and flexible FPGA processors, allowing for more sophisticated trigger algorithms to be applied. The upgraded ECAL will be capable of high-precision energy measurements throughout HL-LHC and will greatly improve the time resolution for photons and electrons above 10 GeV.
\end{abstract}





\section{Introduction}
\label{sec:intro}
The electromagnetic calorimeter (ECAL) is an important component of the CMS detector for the identification and reconstruction of photons and electrons, and is valuable for the measurement of jets and missing transverse momentum. It is composed of a cylindrical barrel (EB), covering the central rapidity region (up to $|\eta|$ = 1.48)  and of two disks called endcaps (EE), which detect incident particles up to $|\eta|$ = 3.0. The detector was designed having in mind the search for the Higgs decay to two photons, for which  excellent energy resolution and efficient identification of photons are required. A precise measurement of energy and momentum of electrons is important for the measurement of Higgs decays into two Z bosons and many other research topics within and beyond the Standard Model.

The ECAL has indeed fulfilled its performance goals with remarkable success. However, it was designed to meet these performance requirements up to an integrated luminosity of 500~fb$^{-1}$. In order to maintain  the current performance up to the integrated luminosity of 4500 fb$^{-1}$ foreseen at the end of the HL-LHC program, an upgrade of the EB readout electronics, which we describe in this report, is needed. The transparency loss in EE will be so significant that the detector will be completely replaced before the start of HL-LHC \cite{hl}. 

\section{The CMS ECAL detector}

The CMS ECAL is a homogeneous calorimeter made of 75848 lead tungstate (PbWO$_4$) scintillating crystals, located inside the CMS superconducting solenoid magnet. The photodetectors are avalanche photo-diodes (APD)
in the barrel and vacuum phototriodes (VPT) in the endcaps. The barrel region is
made of 36 identical supermodules (SM), each containing the crystals, APDs, and
readout electronics. The latter includes very front-end (VFE) cards, which provide pulse amplification, shaping, and digitisation functions, and front-end (FE)
cards which provide data pipeline, data transmission, and trigger primitive formation functions. Electrons and photons are typically reconstructed up to $| \eta | < $ 2.5,
the region covered by the tracker, while jets are reconstructed up to $| \eta |=$ 3.0. The
ECAL energy resolution achieved during 2010 and 2011 ranges from 1.1 to 2.6\% in the barrel and
2.2 to 5\% in the endcaps for photons from Higgs boson decays.

\section{The ECAL upgrade for HL-LHC}

Two of the challenges that the ECAL detector will have to face for the HL-LHC program are aging effects and harsher data-taking conditions. Aging will cause a loss of crystal transparency and an increase in the APD leakage current and therefore noise. To mitigate both effects, the operating temperature will be reduced from the current 18$^\circ$C to 9$^\circ$C. On the other hand, the number of average concurrent interactions per bunch crossing will increase to 200 from the 40 that corresponded to the normal running condition in the first phase of LHC. An improved time resolution will be needed in order to be able to distinguish the primary vertex from the other vertexes in the same bunch crossing. To meet this goal, a faster front-end electronics is being designed. This improvement will also result in  better discriminating power between scintillation light generated by electromagnetic showers from the faster anomalous signals caused by particles hitting the APD directly. The back-end, off-detector electronics is also being redesigned, in order to obtain greater flexibility in the trigger and cope with the increased CMS-wide level-1 (L1) trigger latency (from 3.5 $\mu$s to 12 $\mu$s) and rates (from 100 kHz to 750 kHz) \cite{tdr}.

\section{The upgraded ECAL front-end electronics}
The currently installed  crystals, APDs and motherboards of the EB will not be affected by the upgrade program. The APD output will be sampled and each sample shipped out to the off-detector electronics using high-speed optical links. The front-end will comprise a new, faster analog electronics and the sampling rate will increase from 40 to 160 MHz, with 12-bit resolution, to obtain better time resolution and anomalous signal discrimination. The cost of this improvement is the increase in data rate, for which a loss-less data compression mechanism was devised.

\begin{figure}[htbp]
\centering 
\includegraphics[width=.99\textwidth]{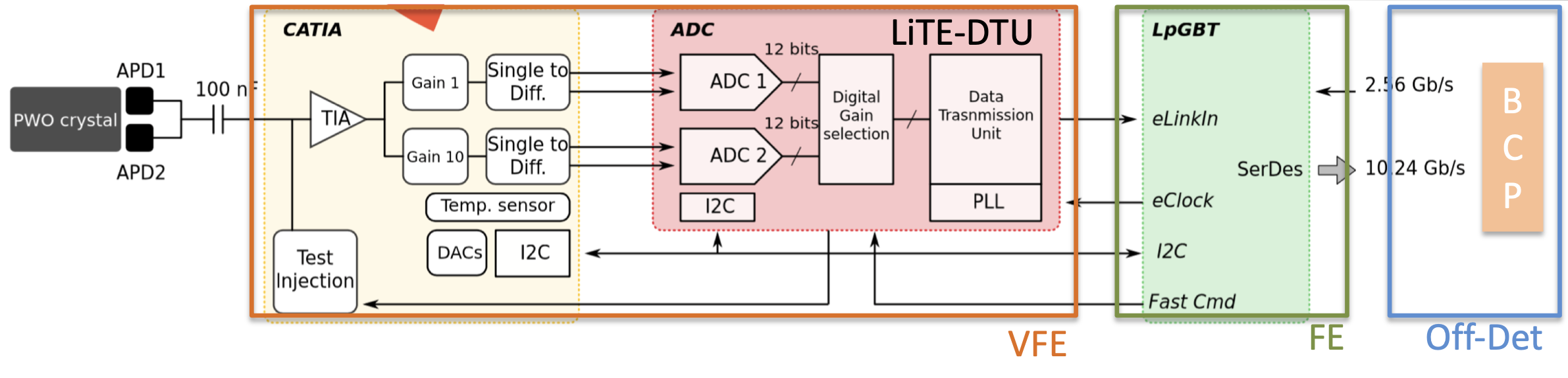}
\caption{Block representation of the EB readout electronics for HL-LHC}\label{scheme}
\end{figure}

Figure \ref{scheme} shows a block diagram of the new readout electronics. The upgraded VFE comprises an analog ASIC called CATIA, which implements two transimpedence amplifiers (TIAs) with 35 MHz bandwidth, and a digital ASIC called LiTE-DTU, comprising two 12-bit ADCs sampling at 160 MHz, data transmission with gain selection and compression.  Each VFE reads five APDs and carries five channels, each channel consisting of a CATIA and a LiTE-DTU. Five VFEs are connected to a single FE card that collects data from the VFEs and transmits to the off-detector electronics using the lpGBT \cite{lpgbt} optical transmission system. Also connected to the FE is a low voltage regulator  board (LVR), a radiation-hard voltage regulator card based on the Feast DC-DC converter. Off-detector, the Barrel Calorimeter Processor (BCP) card will use powerful FPGAs to form the L1 trigger decision and read out the detector.   

\begin{figure}[htbp]
\centering
\includegraphics[width=.49\textwidth]{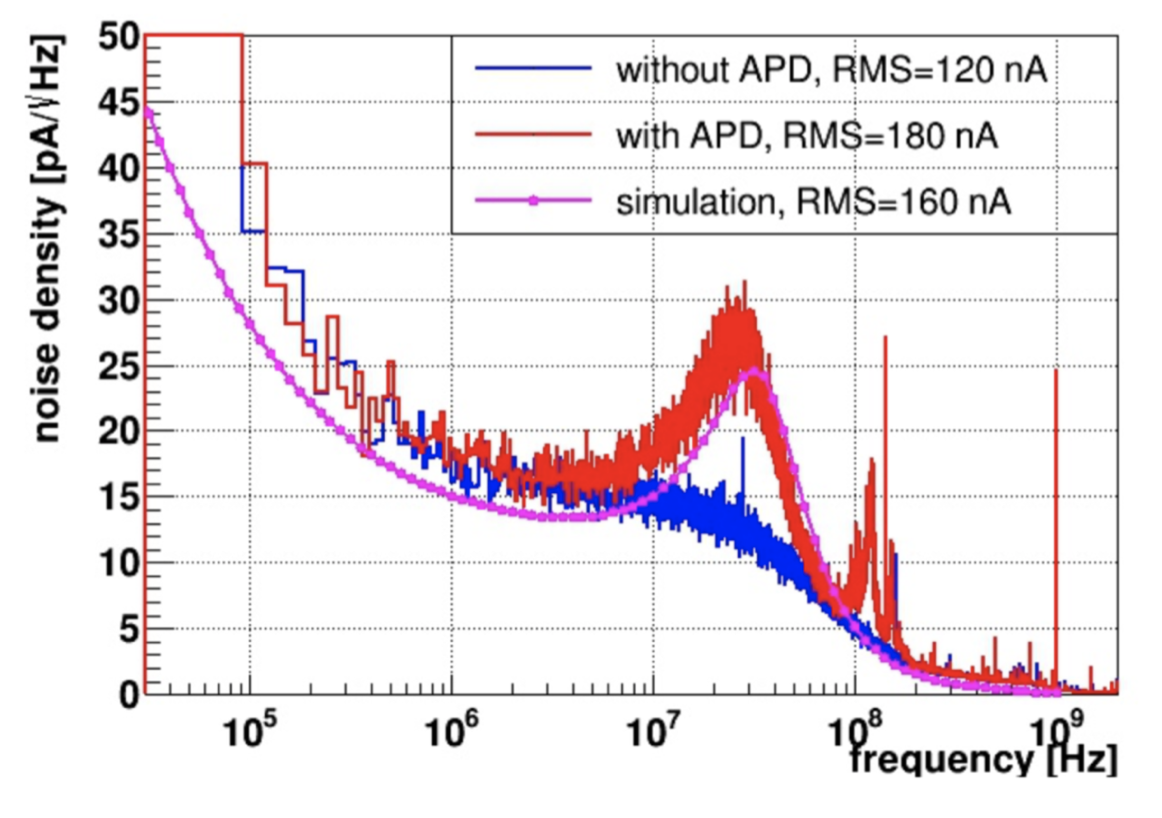}
\includegraphics[width=.49\textwidth]{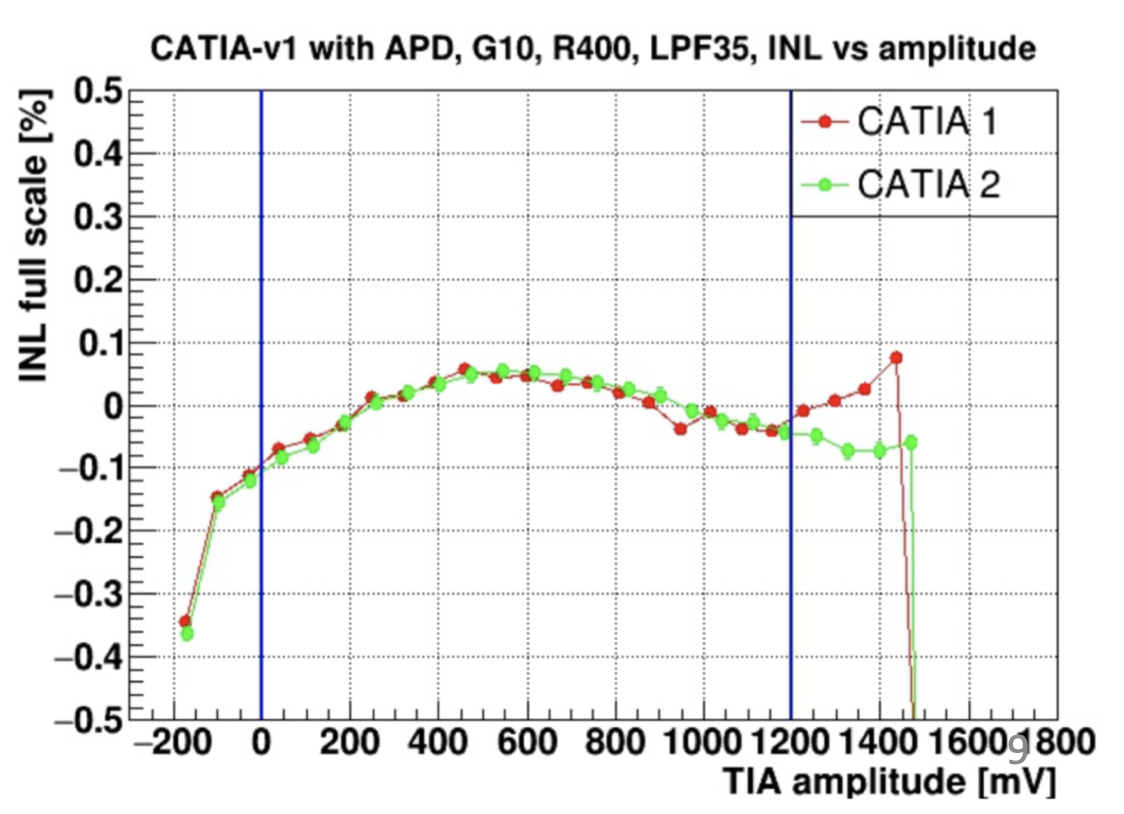}
\caption{\label{figa}Left: noise performance of the CATIA ASIC as a function of frequency, connected and not connected to the APD, compared to simulations. Right: Integral non-linearity measurements of the first and second CATIA prototype, as a function of input signal amplitude. }
\end{figure}

\begin{figure}[htbp]
\includegraphics[width=.55\textwidth]{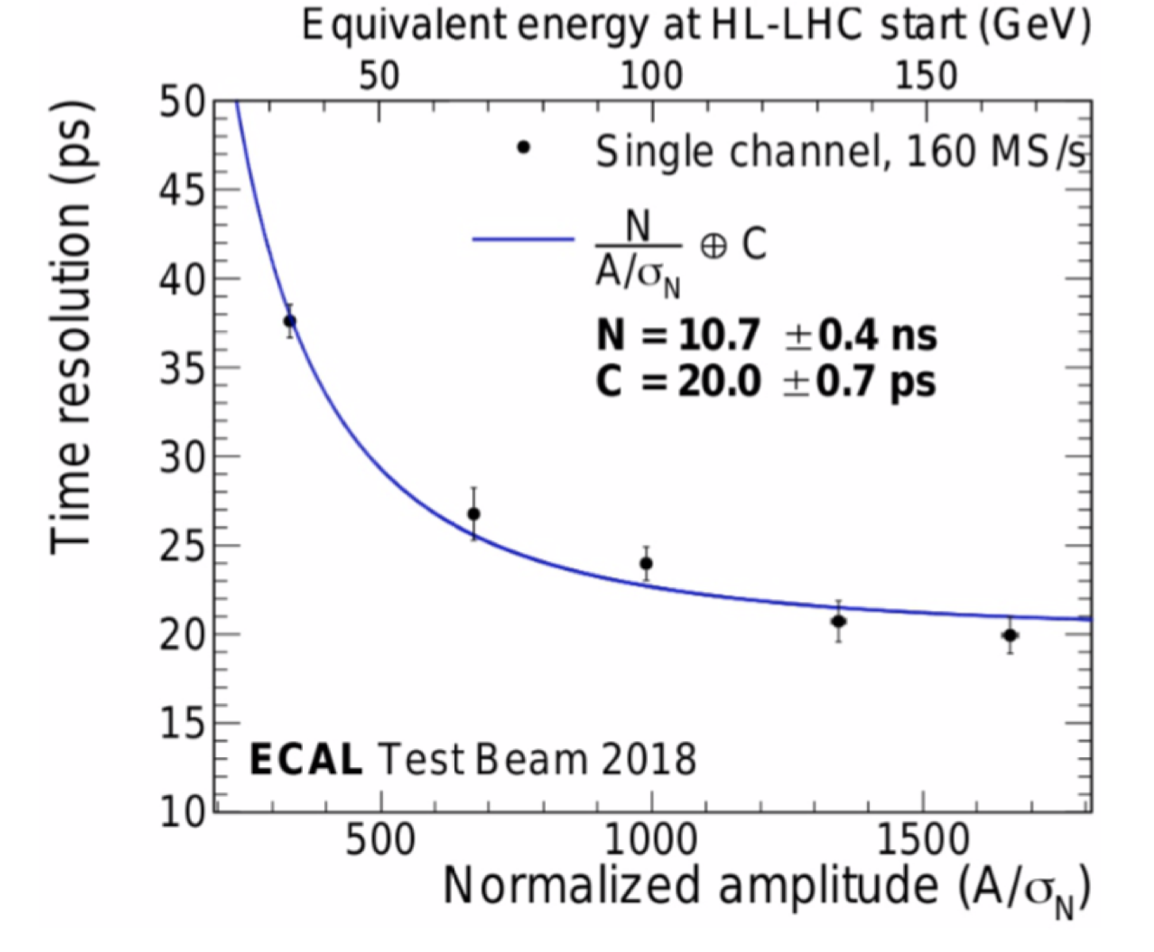}
\includegraphics[width=.35\textwidth, trim = 0  -1.3cm 0 0  0, clip]{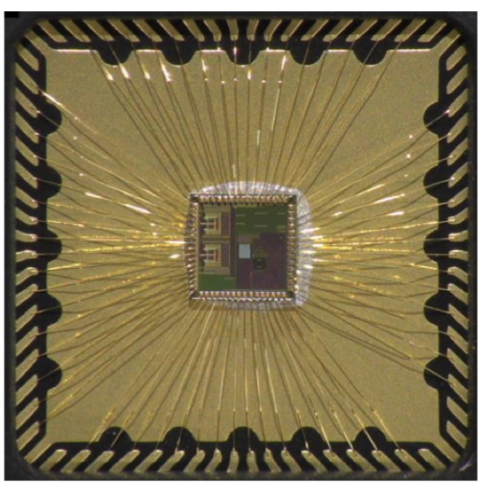}
\caption{\label{figb} Left: time  resolution as a function of normalized amplitude and energy obtained in a test beam campaign with the CATIA ASIC connected to a commercial ADC. Right: the first prototype of the LiTE-DTU ASIC.}
\end{figure}

The CATIA ASIC is built in 130 nm CMOS technology. It comprises a high-gain and a low-gain amplification channel of the APD output, to cope with the dynamic range requirements, and it features test pulse injection. The performances of the CATIA prototype have been tested in test beam campaigns and shown to be  very good in term of noise, linearity, and time resolution. The results are presented in Figs. \ref{figa} and \ref{figb}. Figure \ref{figa}, left, shows the measured noise density spectrum of the CATIA, with and without connection to the APD, compared to simulation. Figure \ref{figa}, right, shows a measurement of the integral non-linearity (INL) of two samples of the CATIA (v1) as a function of the amplitude of an injected test pulse. In the range of interest (0 to 1200 mV) the INL is below 0.1\%. Figure \ref{figb}, left, shows the timing resolution obtained with electrons from a Test Beam at CERN as a function of signal amplitude and energy. A resolution of better than 30 ps is obtained for electrons of energy greater than 50 GeV.

The LiTE-DTU ASIC is built in 65 nm CMOS technology. It includes two 12-bit ADCs running at 160 MHz, designed by a commercial company and specified for a ENOB (equivalent number of bits) of better than 10.2 at 50 MHz. The data transmission unit selects the highest non-saturated gain channel and takes care of data transmission and compression. It includes a PLL circuit for clock generation, it is designed to be TID (total irradiation dose) tolerant up to 20 kGy and it features SEU-tolerant logic. The loss-less data compression mechanism exploits the fact that the spectrum of the hits in each ECAL crystal falls very rapidly with energy. Consequently, the signal is above 2.5 GeV only in a small fraction of events. The compression algorithm was designed to use 6 bits to encode signals below 2.5 GeV, and 12 bits for signals above 2.5 GeV. This allows the reduction of the expected bandwidth occupancy to 1.08 Gb/s, to be compared with the 2.08 Gb/s needed in absence of compression. The first prototype of the LiTE-DTU chip was under test in late 2019. An image of the packaged prototype is visible in Fig. \ref{figb} (right).    

The function of the  FE card is the streaming of digitized data generated on the VFE to the CMS ECAL back-end electronics, system initialization and control of all VFE components, and precise distribution of the clock to VFE cards. It includes four lpGBT downlinks at 10.24 Gb/s and one downlink at 2.56 Gb/s. A prototype is currently being tested with the front-end cards prototypes. 

The off-detector BCP card will be common for ECAL and HCAL, the barrel section of the hadron calorimeter. Its purpose is trigger primitive formation, clock distribution, control, and data readout. It is implemented as an ATCA blade and the V1.0 prototype is based on the Xilinx Kintex Ultrascale KU115 FPGA. The BCP carries a large number of high-speed interfaces to the front-end, the CMS data acquisition, the LHC signaling system, and the neighboring BCPs.

\section{Summary}

The challenging conditions of HL-LHC, with a 4-5 fold increase in occupancy with respect to LHC, require a complete overhaul of the CMS electromagnetic calorimeter electronics. The readout chain has been completely redesigned, while keeping the crystals and photosensors. The bandwidth of the analog readout has been increased by a factor four, as well as the sampling frequency. The trigger hardware was moved off-detector for maximum flexibility. The prototype ASICs and readout boards show excellent performances. At present a vertical integration test, to measure the performance of the readout chain end-to-end, is underway. A larger scale test, with the electronics effectively mounted on a ECAL module, is in preparation. Mass production is foreseen in 2022 and 2023.

\end{document}